\RequirePackage{fix-cm}

\documentclass[smallextended,10pt]{svjour3}       

\usepackage{graphicx} 
\usepackage{natbib}   
\usepackage{mathptmx} 
\usepackage[usenames]{color}
\usepackage[caption=false]{subfig} 

\newcommand{\citeay}[1]{\citeauthor{#1} \citeyear{#1}}
\newcommand{\citepay}[1]{(\citeauthor{#1} \citeyear{#1})}

\journalname{Bulletin of Atmospheric Science and Technology}

\begin{document}

\title{Reforecasting the November 1994 flooding of Piedmont with a convection-permitting model}
\titlerunning{Reforecasting the 1994 Piedmont flood}        
\author{Valerio Capecchi}
\authorrunning{Capecchi V}
\institute{Valerio Capecchi \at
              LaMMA - Laboratorio di Meteorologia e Modellistica Ambientale per lo sviluppo sostenibile\\
              Via Madonna del Piano 10, Sesto Fiorentino, Firenze, Italia\\
              \email{capecchi@lamma.rete.toscana.it}
}
\date{Received: date / Accepted: date}
\maketitle

\begin{abstract}
The Piedmont region in Italy was affected by a heavy rainfall event in November 1994. On the 4th convective cells involved the coastal mountains of the region. On the 5th and early 6th, there were abundant precipitations, related to orographic lift and low-level convergences, in the Alpine area. This study aims to evaluate whether a convection-permitting model provides more valuable information with respect to past numerical experiments.
Results for the 4th of November show that the cloud-resolving model successfully reconstructs the structure of precipitation systems on the downstream side of the coastal mountains. As regards the precipitations of the 5th of November, no added value is found. However, we provide evidence of the anomalously intense transport of moist air from the tropical and subtropical Atlantic and postulate how such transport is responsible for reducing the stability of the flow impinging on the Alps.
\end{abstract}

%
\section{Introduction}\label{sec:intro}
Every year, weather related disasters cause huge damage, significant economic losses and often casualties. According to the ``Atlas of Mortality and Economic Losses from Weather, Climate and Water Extremes 1970-2012'' edited by the World Meteorological Organization, floods and storms are the costliest events in Europe. Among such dramatic cases, the severe weather event of November 1994 in the Piedmont region (north-western Italy) is definitively one to be considered as remarkable. As a consequence of the heavy rainfall, several rivers in the southern part of the region flooded causing huge economic losses (estimated at about 14 billion US dollars) and 70 people died. Past numerical simulations were able to provide quite accurate reconstructions of the event. Nevertheless, the improvements achieved over the last 25 years in weather modelling pose the question of whether we can obtain new insights into the Piedmont flooding by using convective-scale numerical models.
\newline \cite{petroliagis1996} were the first to study the November 1994 Piedmont case (hereinafter P94). They used the ECMWF global model operational at that time, both with the single deterministic forecast and with an ensemble approach. The spectral resolution of their deterministic run (ensemble members) was T213 (T63), which roughly corresponds to about 60 (210) km grid spacing at the Equator. Despite the rough resolution of the model and the fact that the number of members (32 $+$ the control member) of the ensemble was fewer than the one currently used, the authors reached some conclusions about the predictability of the event. Firstly, the day-5 to day-1 T213 forecasts were skilful, with the precipitation patterns well positioned according to observations. Secondly, the day-1 T213 forecast provided a rainfall maximum of 135 $mm$, which, although strongly underestimated with respect to observed data, gave a clear warning of potential severe weather. Finally, the evolution of rainfall predictions in the day-8 to day-3 T63 simulations exhibited consistency, suggesting a significant possibility of an extreme event in the grid point closest to the area of interest. The authors concluded that the probabilistic T63 predictions supported the results of the deterministic T213 forecast and reinforced the degree of confidence that could be associated with it.
\newline Analysing the data registered by automatic weather stations, on the 4th of November intense rainfall rates affected southern Piedmont, in an area between the Maritime Alps and Ligurian Apennines (\citeay{lionetti1996italian}; \citeay{jansa2000western}; \citeay{cassardo2002flood}). \cite{buzzi1998numerical} noted how large errors in their numerical experiments were associated to precipitation forecasts in southern Piedmont, where convection was the main contributor to rainfall. They concluded that these errors were likely due to the lack of an explicit computation of the trajectories of convective cells. A misplacement of precipitation maxima in this area was also found in the work by \cite{ferretti2000numerical}. In another paper, \cite{buzzi2000mesoscale} implemented a high-resolution model simulation of the P94 case with a mesh size of 4 km, but without the explicit computation of convection processes. Also in this case, the authors demonstrated how the precipitation maxima over southern Piedmont were underestimated and misplaced by about 25-30 km to the south of the actual position. As a consequence, the drainage basins of the Tanaro, Bormida and Belbo Rivers in southern Piedmont would not be affected by rainfall runoff with likely errors in hydrological predictions.
\newline \cite{romero1998mesoscale} used the hydrostatic mesoscale model described in \cite{nickerson1986numerical} to analyse the P94 case. Numerical simulations agreed well with observed precipitation patterns, however maximum rainfall predictions were strongly underestimated. By looking at the convective part of the simulated precipitation, the authors argued that the rainfall over Piedmont on the 5th of November had a non-convective nature.
\newline This assessment was confirmed by \cite{doswell1998diagnostic} who analysed satellite images. They concluded that much of the rainfall over the foothills of the Alps in the Piedmont area was associated with upslope flow of weakly stable, moist air. Furthermore, using the ECMWF analysis data, they calculated the Froude number, defined by the relationship
\begin{equation}\label{eq:froude}
Fr=\frac{U}{Nh},
\end{equation}
where $U$ is the horizontal wind speed averaged over the 1000-, 925-, and 850-hPa isobaric levels, $N$ is the stability Brunt-V\"ais\"al\"a frequency calculated from the potential temperature difference between 1000- and 850-hPa levels and $h$ is the height of the orography (which is about 3000 $m$ in the present case). When the Froude number is greater than 1, flow over the orographic obstacle is favoured, thanks to strong low-level winds $U$ or to reduced values of the static stability $N$ (or a combination of the two). On the contrary, the ascent is reduced when $Fr \ll 1$, favouring flow around the obstacle (depending on the geometry of the mountains). \cite{doswell1998diagnostic} found that the Froude number was about 0.90 on the 5th of November. They concluded that in this regime, the uplift flow could be attained only in saturated conditions.
\newline In their idealized numerical study with an L-shape orography (rotated 90 degrees clock-wise), \cite{rotunno2001mechanisms} demonstrated how westward deflection of the southerly sub-saturated flow blocked by the eastern Alps enhanced low-level convergence with the saturated flow directed towards the western Alps. As a consequence, this latter flow was forced to ascend further upstream the mountainous barrier, due to the difference of equivalent potential temperature between the two airflows (see Figure 16 in their paper).
\newline Similar considerations are found in \cite{buzzi1998numerical}. To demonstrate how a relatively stable and moist flux can change the Froude number, they conducted some experiments to assess the role of latent heat exchanges due to condensation and evaporation. A numerical investigation was conducted by suppressing latent heat release due to condensation. Results showed that the low-level flow over the Mediterranean Sea was diverted to the western side of the Alps with respect to the control simulation. As a consequence, the precipitation maxima were found in southern France (see Figures 12 and 13 in their paper). To explain such behaviour, the authors underlined that suppressing the release of latent heat due to condensation, resulted in higher MSLP values in the Alpine region and thus a reduced uplift flow over the mountains. Similar findings and discussions are reported in \cite{ferretti2000numerical}.
\newline Moreover \cite{buzzi1998numerical} and other authors (\citeay{romero1998mesoscale}; \citeay{ferretti2000numerical}; \citeay{jansa2000western}; \citeay{cassardo2002flood}) performed numerical experiments by suppressing (or reducing) the orography of the model grid. All of them found that the precipitation maxima were strongly reduced. Summarising the results in the literature showed that on the 5th of November, the precipitation in the Alps (northern Piedmont) was the result of the combined action of orographic uplift and latent heat exchanges and that convection played a marginal role.
\newline Table \ref{tab:ref} lists past numerical experiments dealing with the P94 case. All the models deployed some kind of parametrization scheme for convection processes. Mesh sizes ranged between 60 km for the global model used in \cite{petroliagis1996} up to 4 km, which is the resolution of the experiments by \cite{buzzi2000mesoscale}.
\begin{table}
\caption{Past numerical investigations on the November 1994 Piedmont case}\label{tab:ref}
\begin{tabular}{lllll}
\hline\noalign{\smallskip}
Reference & Model & Resolution & Vertical & Forecast \\
          &       & (grid spacing) & levels & length \\
\noalign{\smallskip}\hline\noalign{\smallskip}
\cite{petroliagis1996} & IFS (EPS) & T213 (T63) & 31 (19) & 120 ours\\
\cite{romero1998mesoscale} & SALSA & 20 km & 30 & 30 hours\\
\cite{buzzi1998numerical} & BOLAM & 30 and 10 km & 36 & 36 hours\\
\cite{buzzi2000mesoscale} & BOLAM & up to 4 km & 40 & 30 hours\\
\cite{jansa2000western} & HIRLAM & 40 km & 31 & 48 hours\\
\cite{ferretti2000numerical} & MM5 & 30 and 10 km & 23 & 48 hours\\
\cite{cassardo2002flood} & RAMS & 15 km & 35 & 84 hours\\
\noalign{\smallskip}\hline
\end{tabular}
\end{table}

An analysis of the P94 case can be conducted in light of an atmospheric river (AR) landfall. Since the seminal work by \cite{zhu1998proposed}, AR theory and field experiments has received much attention in relatively recent years (see for instance \citeay{ralph2004satellite}, \citeay{lavers2013nexus} and \citeay{krichak2016discussing}). Roughly speaking, ARs are defined as narrow tongues of moist air in the lower troposphere responsible for the transport of tropical water vapour into the extratropics. Although ARs are mainly studied for their impacts on the western coasts of the United States, some papers demonstrated that heavy precipitation and flood events in Europe are often linked to ARs landfall (\citeay{stohl2008remote}, \citeay{lavers2013nexus}, \citeay{krichak2016discussing}). Furthermore, as stated in \cite{krichak2016discussing}, the regions mostly prone to the impacts of ARs landfall are mountainous, providing the necessary uplift for significant rainfall. 
\newline Over the last few decades, convective-scale numerical weather prediction modelling improved tremendously \citepay{sun2014use} and is now facing new challenges \citepay{yano2018scientific}. Nevertheless the predictability of small-scale, high-impact weather remains limited due to (i) approximations in the reconstruction of fine-scale processes \citepay{leutbecher2017stochastic}, and (ii) the chaotic nature of the weather system causing small errors in the initial conditions to grow rapidly \citepay{palmer2001nonlinear}. Nowadays, technological capability allows operational convection-permitting models to be run with horizontal mesh sizes less than 2 km as well as ensemble systems with horizontal mesh size about 3 km. To mention a few examples in national weather services in Europe, the mesh size of the operational model of M\'et\'eo-France (AROME) is about 1.3 km; the UK Met Office runs a 1.5 km grid length model four times per day and has plans to deploy a sub-kilometre scale version in the near future. The newest version of the ICON model at the German DWD weather service is a seamless model that can be run with a mesh size of less than 1 km. As regards probabilistic prediction systems, examples include COSMO-DE at 2.8 km horizontal resolution \citepay{peralta2012accounting}, MOGREPS-UK at 2.2 km horizontal resolution \citepay{hagelin2017met} and AROME-EPS at 2.5 horizontal resolution \citepay{raynaud2015comparison}.
\par The goal of this paper is to perform a reforecast of the P94 case. We present a numerical reconstruction by applying cutting-edge regional convection-permitting NWP modelling, fed by recently updated global analyses. The purpose is twofold: as regards the rainfall observed on the 4th of November, the challenge is to improve the prediction of the convective precipitations in southern Piedmont, where the majority of damage and casualties occurred. As regards the rainfall observed on the 5th of November, we provide evidence of an AR landfall during the days of the P94 case.
In addition, we investigate if a better quantitative precipitation forecast can be accomplished by deploying a high-resolution model, which uses an accurate description of topography.
\par We conclude this introduction by stressing the fact that evaluating whether and how current state-of-the-art numerical models can simulate past high-impact events is essential to  understand the information content of current operational forecasting systems in order to predict future extreme events. 

\section{Synoptic overview}
\begin{figure}
  \includegraphics[width=0.75\textwidth,angle=-90]{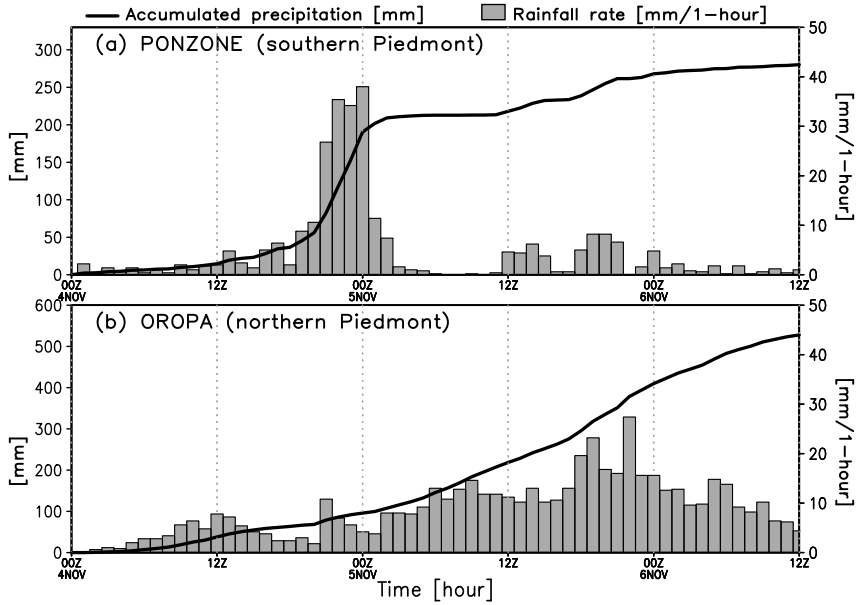}
  \caption{Accumulated precipitations recorded in the period from 00 UTC 4th November 1994 to 12 UTC 6th November 1994 by the rain-gauges at Ponzone and Oropa, whose geographical positions are shown in Figure \ref{fig:domain}}\label{fig:pcp}
\end{figure}

Although detailed synoptic descriptions of the P94 case can be found in the papers listed in Table \ref{tab:ref}, we include here a short summary to make this paper self-contained. October 1994 was characterised by a remarkable weather variability over north-western Italy. Accumulated rainfall recorded by automatic weather stations was above average in several districts of Piedmont.
Between the 2nd and 4th of November, there was an upper level cyclonic circulation over the northern Atlantic area \citepay{petroliagis1996}. ECMWF analysis data valid at 12 UTC of the 4th of November showed a trough with an axis extending from the British Isles to the Iberian peninsula (see Figure 3b in \citeay{buzzi1998numerical}). This configuration activated the advection of warm and moist air over the western/central Mediterranean Sea towards southern France and northern Italy. As recorded by rain-gauges data \citepay{lionetti1996italian}, this time practically coincided with the beginning of the heavy precipitations in southern Piedmont. For instance, see in Figure \ref{fig:pcp}a the accumulated precipitation recorded at the Ponzone rain-gauge (whose location is indicated in Figure \ref{fig:domain}).
One day later, the movement of the trough eastwards was slow, due to the ridge present over the central/eastern Mediterranean Sea (see Figures 4b in \citeay{buzzi1998numerical} and 1a in \citeay{petroliagis1996}). As a consequence, a double front-like structure formed west of Italy as confirmed by numerical simulations and Meteosat satellite images (\citeay{buzzi1998numerical}; \citeay{doswell1998diagnostic}). This configuration is not uncommon in the western Mediterranean and is often associated, as in the P94 case, with intense precipitations in the Alpine region; see the plots in Figure 4 of \cite{lionetti1996italian} and the accumulated precipitation recorded at Oropa rain-gauge in Figure \ref{fig:pcp}b. We note how rainfall rates at Oropa reached 27 $mm$ hour$^{-1}$ with an average value on the 5th of November about 10 $mm$ hour$^{-1}$. Instead, data recorded at Ponzone (Figure \ref{fig:pcp}a) reported a maximum value of 38 $mm$ hour$^{-1}$ with about 135 $mm$ in the last four hours of the 4th of November. Thus the profiles of precipitation rates of Ponzone rain-gauge  (southern Piedmont) exhibited much higher time variability, indicating the presence of convection. This assessment is also found in  \cite{buzzi1998numerical}.
\newline We conclude this brief overview of the P94 case by stressing the fact that, as demonstrated by the numerical experiments conducted by \cite{romero1998mesoscale} and \cite{cassardo2002flood}, the contribution to the precipitations of the Mediterranean sea surface evaporation was likely not important.

\section{Model and methods}\label{sec:data}

\subsection{Model description}
\begin{figure}
  \includegraphics[width=0.75\textwidth,angle=90]{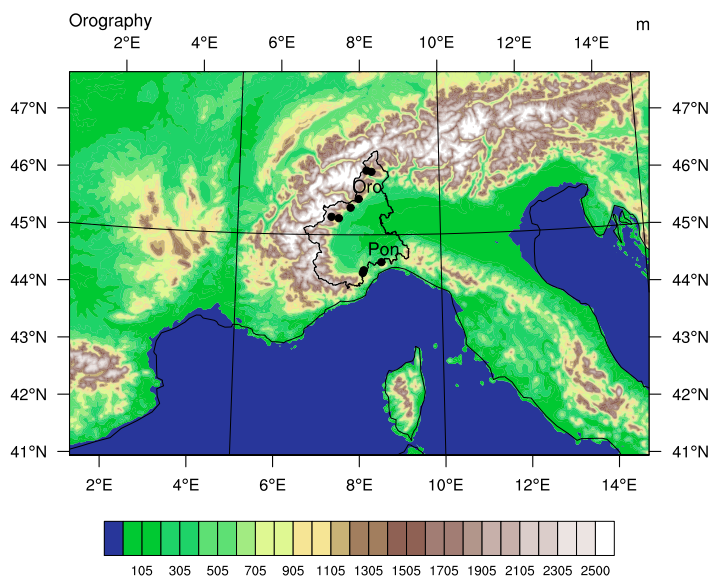}
  \caption{Orography of the domain of the Meso-NH simulation. Horizontal grid spacing $\Delta x$ is 2.5 km. The locations of nine rain-gauges are indicated by black dots: three in southern Piedmont and six in the north of the region, at the foothills of the Alps. The accumulated rainfall recorded at Oropa (``Oro'') and Ponzone (``Pon'') are shown in Figure \ref{fig:pcp}. The political borders of the Piedmont region are indicated by the black line}\label{fig:domain}
\end{figure}

In this paper, we present high-resolution simulations of the P94 case performed using the Meso-NH model. This is a French research community model, jointly developed by the Centre National des Recherches M\'et\'eorologiques (CNRM) and Laboratoire d'A\'erologie (LA) at the Universit\'e Paul Sabatier (Toulouse). It is designed to simulate the time evolution of several atmospheric variables ranging from the large meso-$\alpha$ scale ($\simeq2000 km$) down to the micro-$\gamma$ scale ($\simeq20 m$), typical of the Large Eddy Simulation (LES) models. For a general overview of the Meso-NH model and its applications see \cite{lafore1997meso} and \cite{lac2018overview}, while the scientific documentation is available on the model's website. For this study, we used the version 5.4.1 (released in July 2018). The geographical extent of the simulations is shown in Figure \ref{fig:domain}; no grid-nesting was implemented. As regards microphysics, we set the one-momentum ICE3 scheme \citepay{caniaux1994numerical}, that takes into account five water species (cloud droplets, raindrops, pristine ice crystals, snow or aggregates, and graupel). The convection, both deep and shallow, was explicitly resolved. The Runge-Kutta centred 4th order scheme was chosen for momentum advection. This scheme is recommended when using, as in the present paper, the CENT4TH (4th order CENtred on space and time) advection scheme. The CENT4TH was chosen because of its numerical stability, although it is more time-consuming than other options \citepay{lunet2017combination}. Some specific parameters of the model for this study are summarised in Table \ref{tab:settings}.
\begin{table}
\caption{Key parameters of the Meso-NH model settings}\label{tab:settings}
\begin{tabular}{l l}
\hline\noalign{\smallskip}
\textbf{Variable} & \textbf{Value} \\
\noalign{\smallskip}\hline\noalign{\smallskip}
Rows$\times$Columns & 300$\times$450\\
Horizontal grid spacing & 2.5 km\\
Vertical resolution & 52 levels (up to 20 km)\\
Time step &  10 s \\
Cumulus convection &  explicit (no parametrization)\\
Microphysics option &  ICE3 scheme \citepay{caniaux1994numerical} \\
\noalign{\smallskip}\hline
\end{tabular}
\end{table}

To drive the Meso-NH simulations, initial and boundary conditions were produced using a recent version of the ECMWF-IFS model (cycle 41r2, the same used to produce the ERA5 data). The spectral resolution of the model is TL1279, which corresponds to about 16 km grid spacing. Boundary conditions are provided every 3 hours. 

\subsection{Methods}
As proposed in the seminal work by \cite{zhu1998proposed}, ARs can be detected by looking at the vertically integrated horizontal water vapour transport (hereafter, integrated vapour transport, IVT) defined as
\begin{displaymath}
IVT=\sqrt{\left(\frac{1}{g}\int_{p_0}^{p_{top}} qu\ dp\right)^2+\left(\frac{1}{g}\int_{p_0}^{p_{top}} qv\ dp\right)^2}
\end{displaymath}
where $q(p)$ is the specific humidity in ${Kg}{Kg}^{-1}$, $u(p)$ and $v(p)$ are the zonal and meridional components respectively of the horizontal wind vector in $ms^{-1}$, $g$ is the acceleration due to gravity, $p_0$ and $p_{top}$ are the 1000- and 300-hPa isobaric levels respectively. The algorithms based on this approach \citepay{rutz2014climatological} declare grid points as interested by an AR if they satisfy the condition that IVT exceeds a predefined threshold, which is normally equal to $250\ kg\ m^{-1} s^{-1}$. \cite{ralph2019scale} proposed a scale to characterize the strength and impacts of ARs. It is based on the analysis of both the maximum value of IVT at a given location during the AR event (i.e. $IVT \ge 250\ kg\ m^{-1} s^{-1}$) and the AR duration. The IVT maps presented below were calculated using the ERA5 data \citepay{p19027}.

\section{Results}\label{sec:res}
\begin{figure}
  \includegraphics[width=0.75\textwidth,angle=-90]{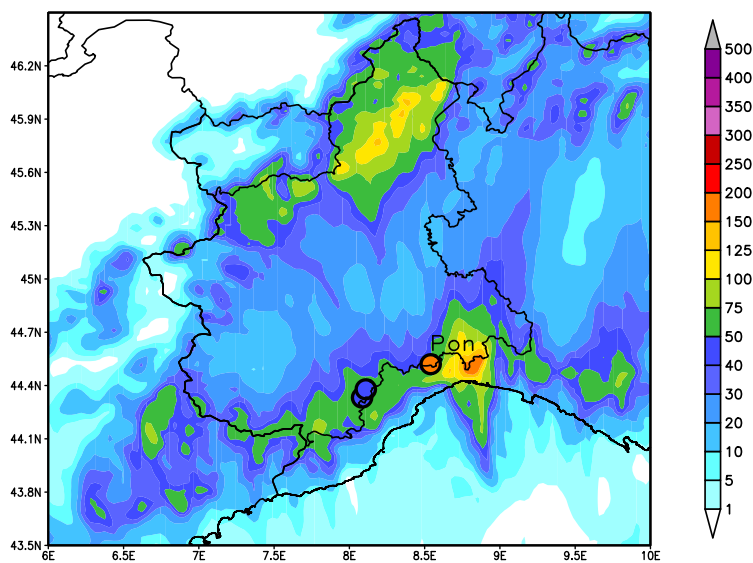}
  \caption{24-hour rainfall forecast of the Meso-NH model initialised on the 4th of November at 00 UTC}\label{fig:MNH:P1104}
\end{figure}

\begin{figure}[ht]%
 \centering
 \subfloat[Transect of the vertical cross section]{\includegraphics[width=0.5\textwidth,angle=0]{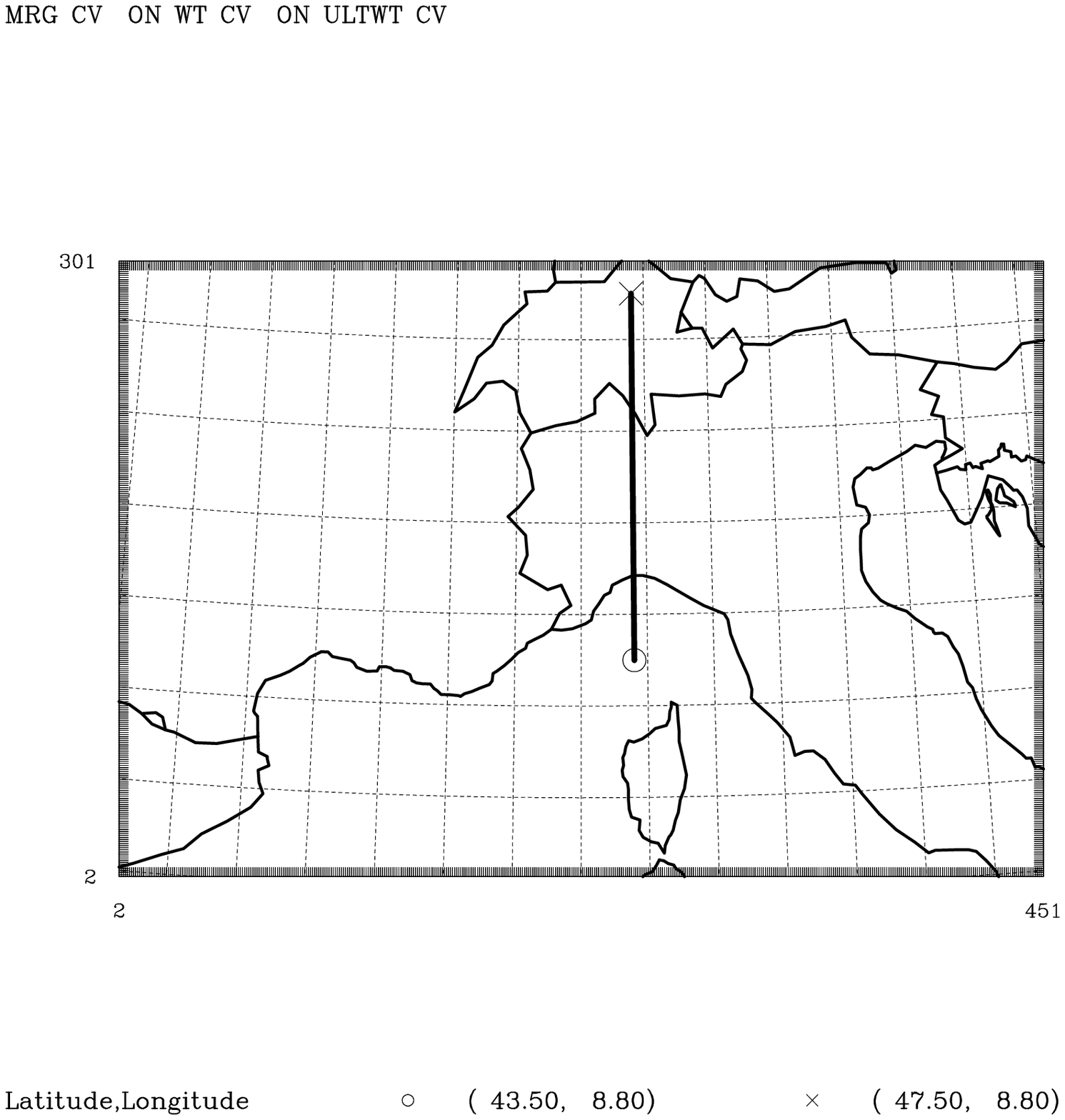}\label{fig:transect}}~~
 \subfloat[Mixing ratio for cloud]{\includegraphics[width=0.5\textwidth,angle=0]{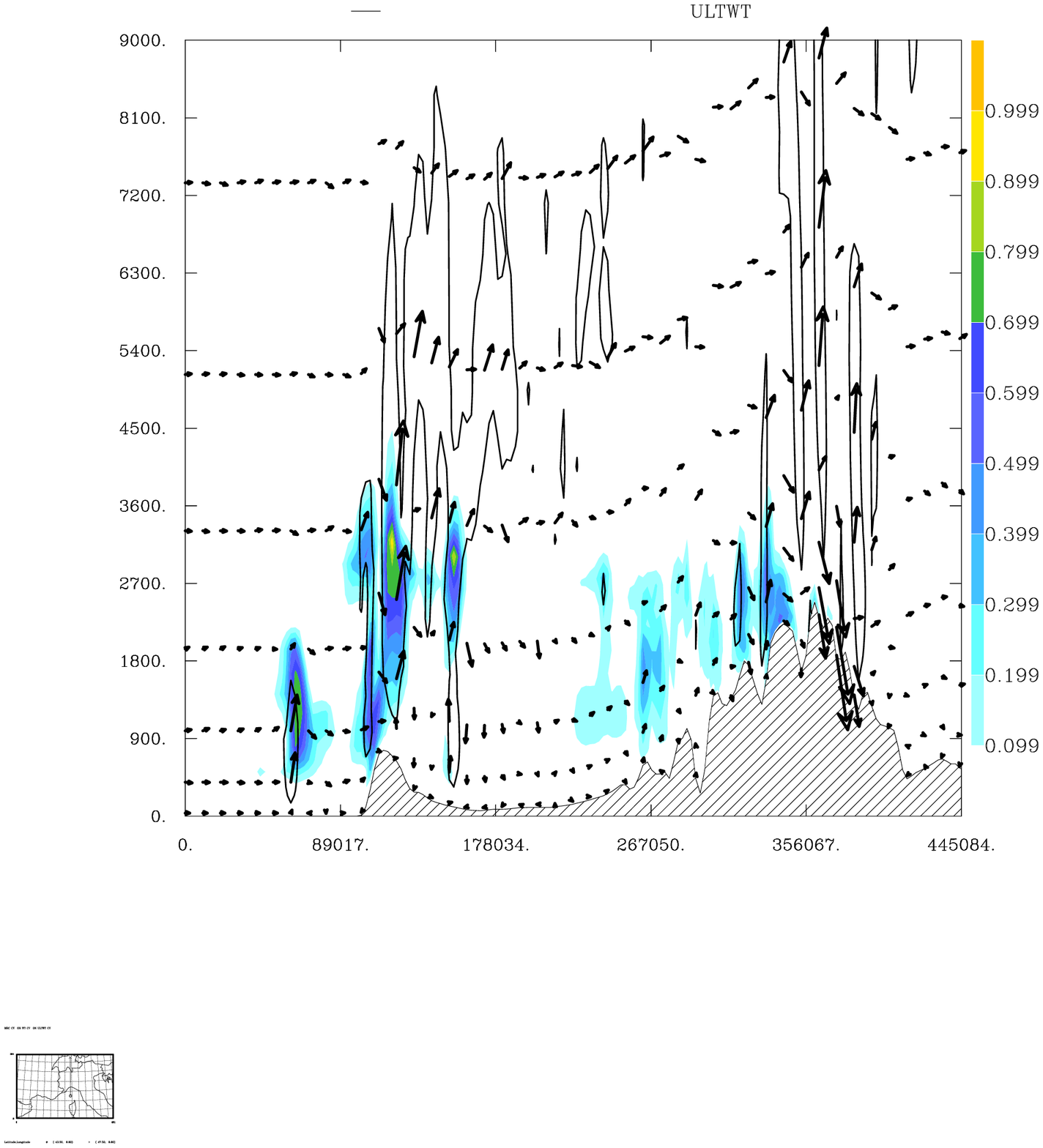}\label{fig:MRC}}\\
 \subfloat[Mixing ratio for rain]{\includegraphics[width=0.5\textwidth,angle=0]{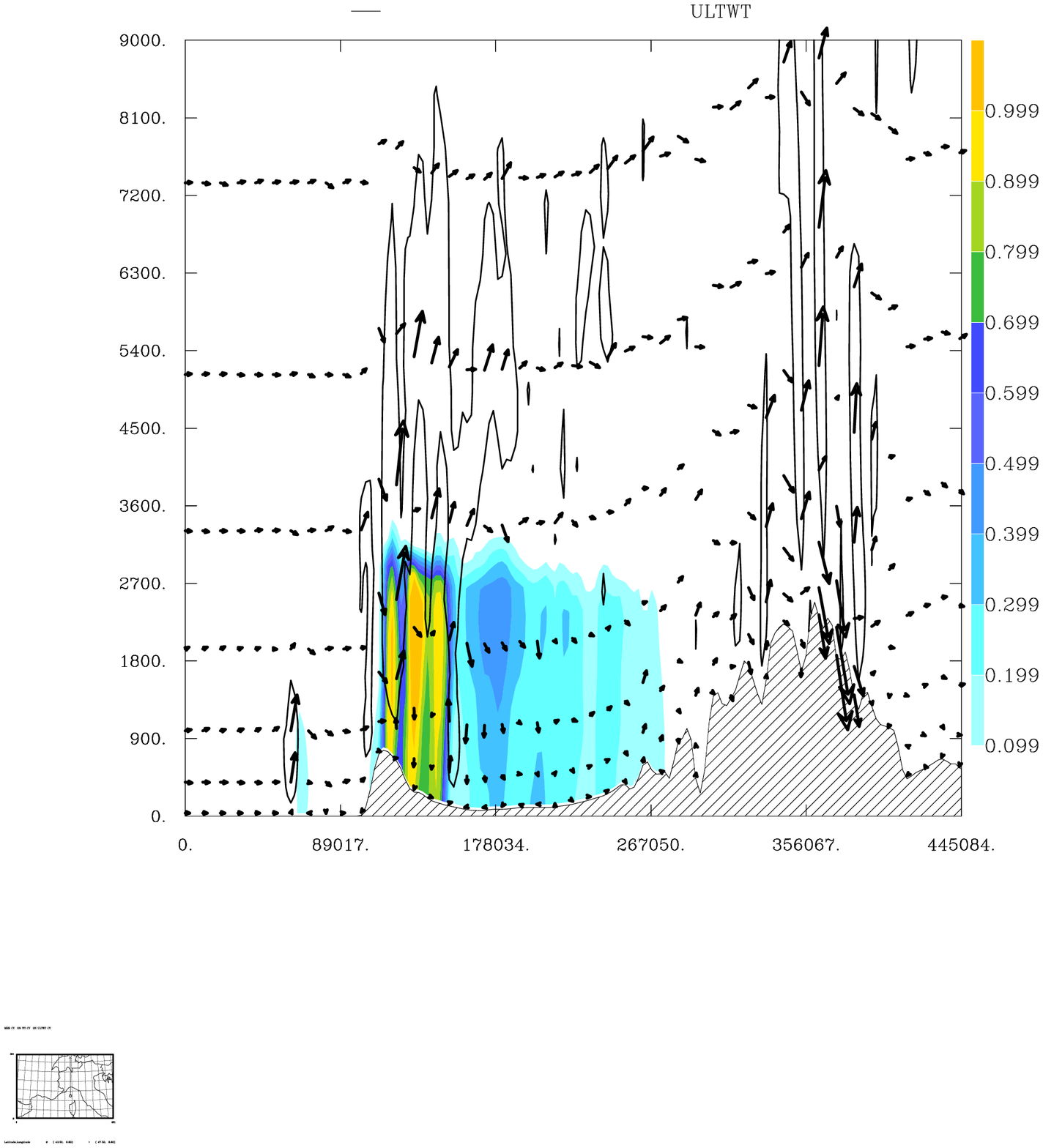}\label{fig:MRR}}~~
 \subfloat[Mixing ratio for graupel]{\includegraphics[width=0.5\textwidth,angle=0]{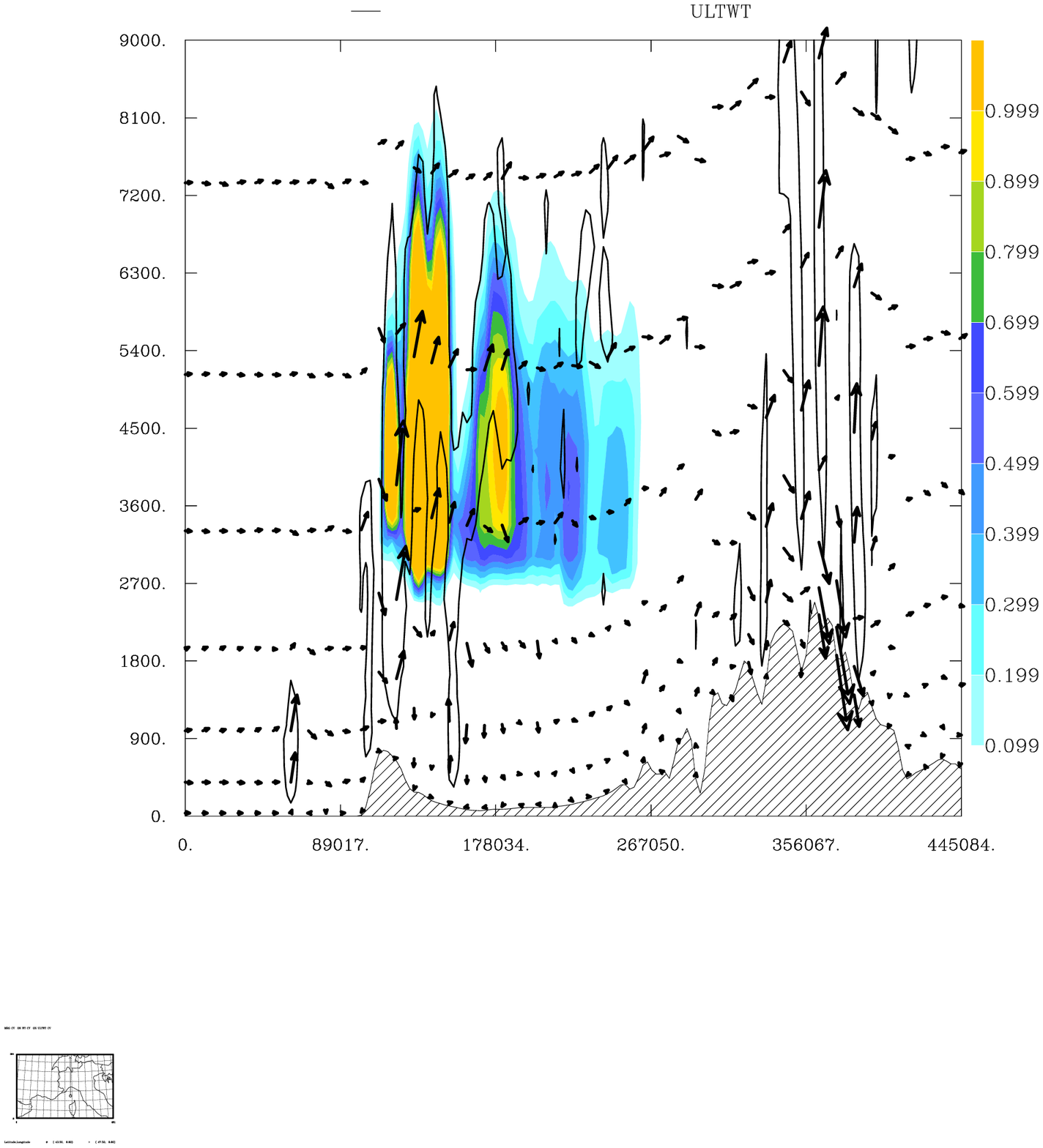}\label{fig:MRG}}
 \caption{Vertical cross section of the Meso-NH forecast valid at 13 UTC of the 4th of November. In panels (b)-(d), wind vectors overlap the mixing ratios of three classes of hydrometeors: (b) cloud water, (c) rain and (d) graupel. Contour intervals indicate vertical velocities greater than 1 $ms^{-1}$. The transect is in panel (a). On the X-axis of panels (b)-(d) is indicated the distance (in $m$) from the starting point ($\circ$ symbol in panel (a)) of the transect whereas on the Y-axis the altitude in $m$ is indicated. The model orography is dashed}
 \label{fig:MNH:P1104:CV}%
\end{figure}

In Figure \ref{fig:MNH:P1104} we show the 24-hour accumulated precipitations predicted by the Meso-NH model for the 4th of November. Observed amounts (coloured circles) overlap the forecasts data. Starting time of the simulation is 00 UTC 4 November 1994. Forecasts initialised one and two days in advance do not provide substantial differences to the one shown here. From a visual inspection of Figure \ref{fig:MNH:P1104} the pattern of precipitation forecast is in good agreement with observed values. However, the simulation tends to slightly overestimate rainfall along the coastal mountains (that is in southern Piedmont). As regards the comparison with observations collected at Ponzone rain-gauge, the maximum rainfall is misplaced by about 20 km.
In Figure \ref{fig:MNH:P1104:CV} we show the vertical cross section at 13 UTC of the 4th of November of a transect (indicated in panel \ref{fig:transect}) close to the Ponzone rain-gauge. The panels \ref{fig:MRC}, \ref{fig:MRR} and \ref{fig:MRG} show the mixing ratio of cloud droplets, rain and graupel respectively and the wind vectors at model levels. Convection is visible on the lee side of the coastal mountains associated to moderate to strong vertical velocities. The model reconstructs a vertical structure of clouds and hydrometeors up to mid-troposphere and above. 

\begin{figure}
  \includegraphics[width=0.75\textwidth,angle=-90]{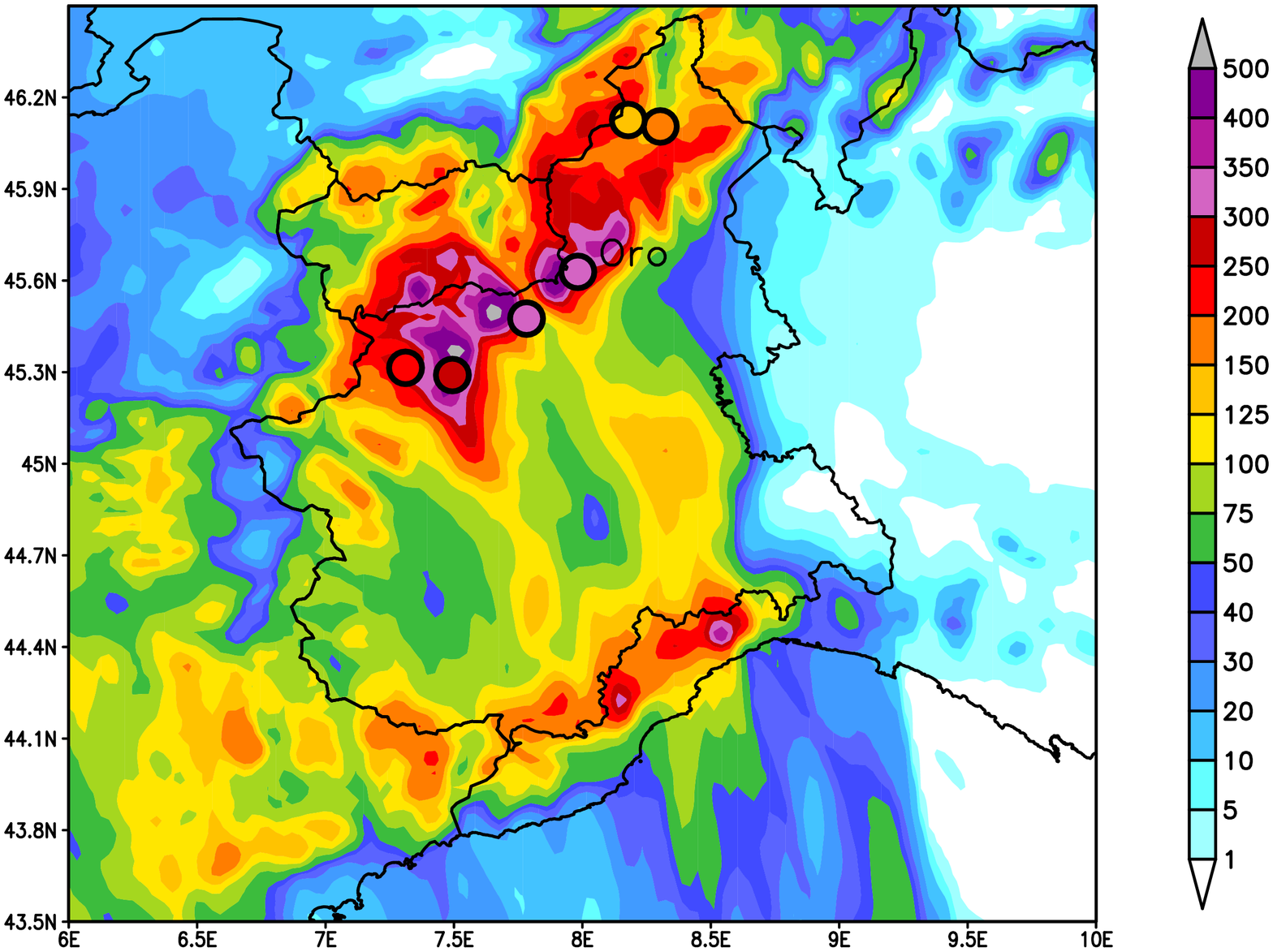}
  \caption{24-hour rainfall forecast of the Meso-NH model initialised on the 5th of November at 00 UTC}\label{fig:MNH:P1105}
\end{figure}

In Figure \ref{fig:MNH:P1105} we show the 24-hour accumulated precipitations predicted by the Meso-NH model for the 5th of November. Starting time of the simulation is 00 UTC 5 November 1994. Forecasts initialised one and two days in advance do not provide any substantial difference to the rainfall map presented here. We compared the 24-hour rainfall amounts recorded by the six rain-gauges located in northern Piedmont (shown in Figure \ref{fig:MNH:P1105} with the coloured circles) with the prediction in the grid points closest to each rain-gauge. We found that the mean error, often referred to as additive bias, is about 27 $mm$, whereas the root mean square error is about 85 $mm$.

We note that maximum rainfall forecasts on the Alps seem to be unrealistic with values up to 500 $mm$ or more in correspondence with mountain peaks.
\par
\begin{figure}
  \includegraphics[width=0.75\textwidth,angle=-90]{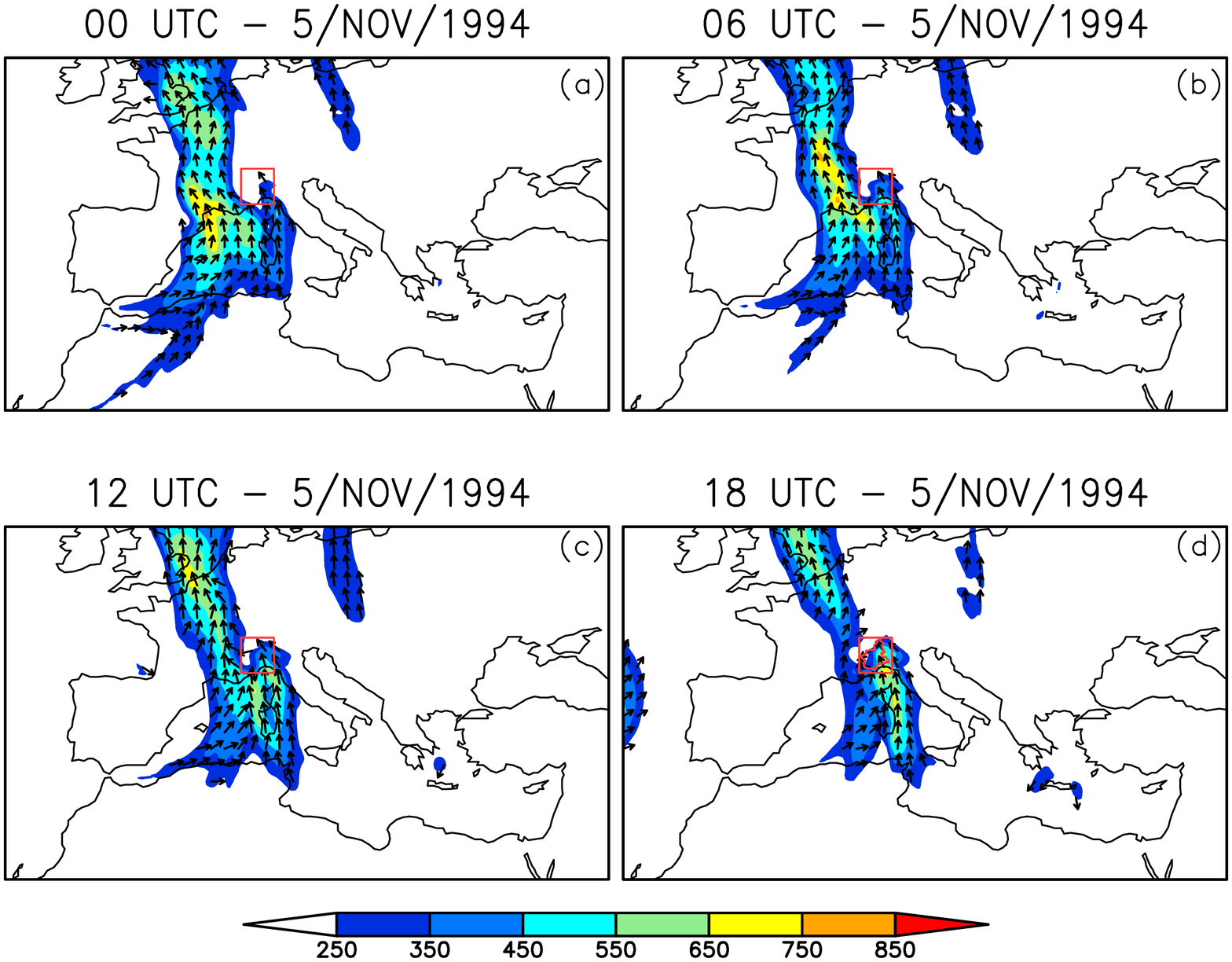}
  \caption{Vertically integrated horizontal water vapour transport (IVT) values expressed in $kg\ m^{-1} s^{-1}$ for the four 6-hour time steps on the 5th of November 1994. Wind vectors at 850-hPa isobaric level are plotted only where $IVT \ge 250\ kg\ m^{-1} s^{-1}$. ERA5 data were used to produce the plots. The Piedmont region is marked with the red box}\label{fig:AR1}
\end{figure}

\begin{figure}
  \includegraphics[width=0.75\textwidth,angle=-90]{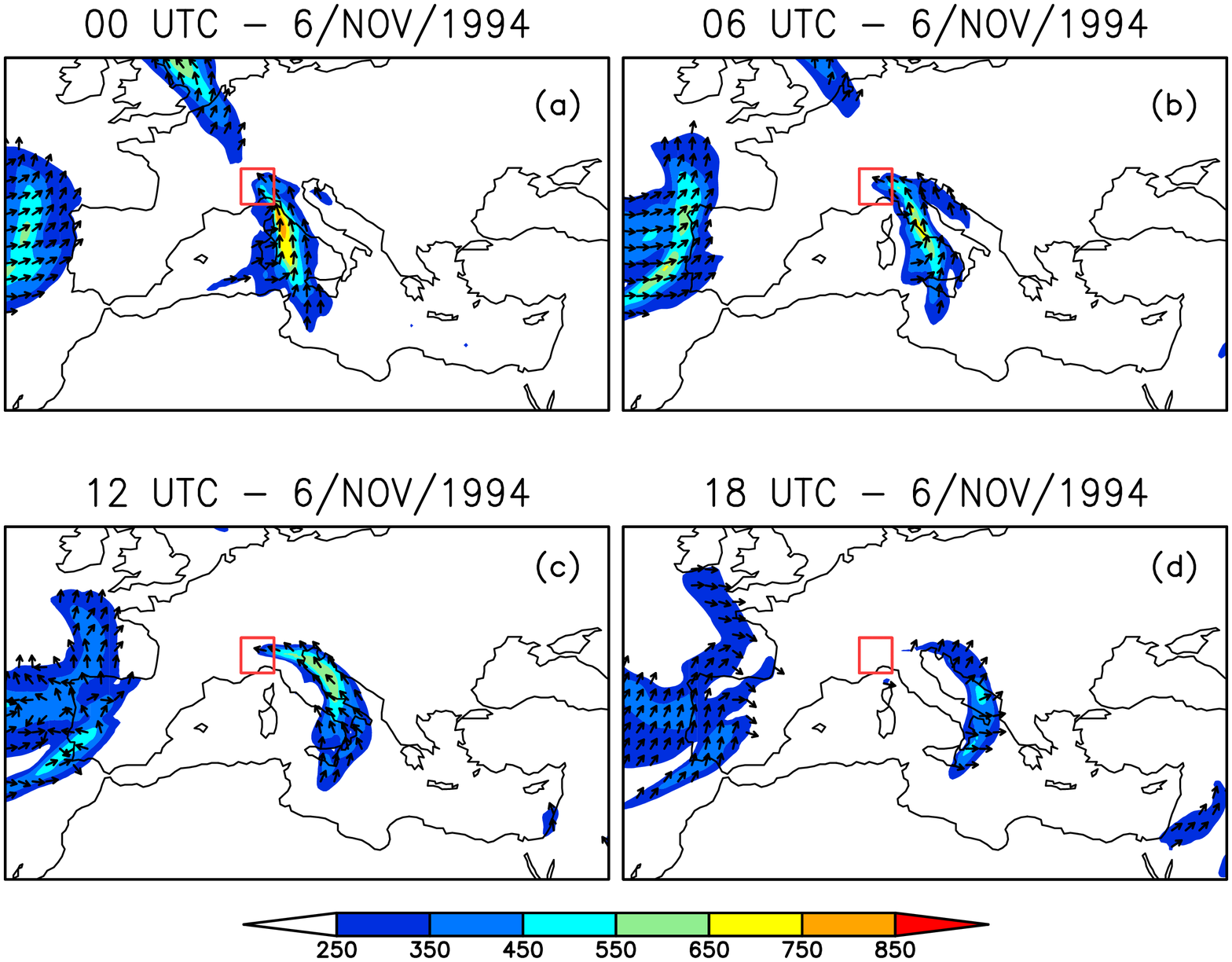}
  \caption{As Figure \ref{fig:AR1} but for the 6th of November 1994}\label{fig:AR2}
\end{figure}

 In Figures \ref{fig:AR1} and \ref{fig:AR2}, we show the IVT maps for the four 6-hour steps of the 5th and 6th of November respectively. On the 5th of November (Figure \ref{fig:AR1}), an elongated band of moisture transport connects the coasts of northern Africa to the British Isles, involving part of the Italian peninsula and France. This band is about 2100 km long and 500 km wide. Elevated IVT values, greater than the critical threshold of $250\ kg\ m^{-1} s^{-1}$, are found in the area of interest, which is indicated in the plots with a red rectangle. On the 6th of November, this moist corridor is split in two parts: one to the north of the Alps and one to the south. However, we can still observe high values of IVT ($\ge 250\ kg\ m^{-1} s^{-1}$) in the north-eastern part of the Piedmont area, at least in the early part of the day (i.e. until 06 UTC). Following the intensity scale proposed by \cite{ralph2019scale}, we note that for the Piedmont region this yields the following: (i) AR conditions (i.e. $IVT \ge 250\ kg\ m^{-1} s^{-1}$) lasted for at least 24 hours (see the maps in Figures \ref{fig:AR1}b,c,d and \ref{fig:AR2}a) and (ii) maximum instantaneous IVT values exceeded 600 $kg\ m^{-1} s^{-1}$ on the 5th of November (plot not shown). We can thus classify the AR that involved the P94 case as category 2 out of 5 categories following \cite{ralph2019scale}.

\section{Discussions and conclusions}\label{sec:dis}
To mark the twenty-fifth anniversary of the Piedmont flooding, we revisited the P94 case by applying cutting-edge regional NWP modelling. High-resolution simulations were forced by initial and boundary conditions produced \textit{ad-hoc} with a recent version of the ECMWF-IFS global model. The goal of the work was twofold: firstly, we wanted to investigate whether a convection-permitting model is able to reconstruct the pre-frontal convection activity observed on the 4th of November in southern Piedmont. Secondly, as regards the orographic precipitation on the 5th of November in the Alpine area, we wanted to assess any new insights of a high-resolution simulation. In fact, previous studies used numerical weather models, often hydrostatic, with the parametrization of convection processes and mesh sizes in the order of few tens of kilometres.
\par As regards the convective precipitation in southern Piedmont on the 4th of November, the Meso-NH forecast reconstructs the dynamics of the event well, apart from the fact that the maximum value seems to be misplaced by a few tens of kilometres to the east of the actual position. What is important to stress is that convective cells are reconstructed in the lee side of coastal mountains in southern Piedmont, as can be appreciated by looking at the vertical cross sections in Figure \ref{fig:MNH:P1104:CV}. Using a convection-permitting model, as suggested by \cite{buzzi1998numerical}, was likely the key to achieving such a result. This is an improvement with respect to previous studies (\citeay{buzzi1998numerical}; \citeay{romero1998mesoscale}; \citeay{ferretti2000numerical}; \citeay{cassardo2002flood}), which lacked the reconstruction of precipitation patterns on both the north and south sides of the coastal mountains.
\par As regards the orographic precipitation in northern Piedmont on the 5th of November, the high-resolution and convection-permitting Meso-NH forecast does not add any new insights with respect to past numerical investigations. As found in previous studies, there is a fairly good consistency between model predictions and rain-gauges data, although the statistical analysis shown here might not be representative due to the scarce number of observations used. Large amounts of rainfall are predicted in the Alps in correspondence to mountain peaks, which seem to be unrealistic given the observations available (see data in \citeay{lionetti1996italian} and \citeay{cassardo2002flood}). The overestimation of the model might be due to the implementation of the ICE3 microphysics scheme, which produces an excess of precipitating graupels or other hydrometeors. \cite{rotunno2001mechanisms} suggested that the simpler Kessler scheme \citepay{kessler1969distribution} contains the essential set of water categories relevant for the P94 case. This scheme takes into account three type of hydrometeors: vapour, cloud water and rain. However, a test Meso-NH simulation using the Kessler scheme provided results that do not differ very much from the ones presented here (map not shown). We give the results with the ICE3 scheme, because it is the most commonly used one-moment scheme in Meso-NH \citepay{lac2018overview} and thus the results obtained can be compared with similar works or case studies. Moreover it is more appropriate to describe the hydrometeors involved in the convective precipitation on the 4th of November as demonstrated by the vertical cross sections shown in Figure \ref{fig:MNH:P1104:CV}.
\newline We conclude the discussion on the Meso-NH simulations by speculating why longer forecasts than the ones presented here provide results that do not differ substantially from those shown in Figures \ref{fig:MNH:P1104} and \ref{fig:MNH:P1105}. We guess that one of the reasons lies in the improvements gained over the last 25 years by global NWP modelling. Indeed, in the paper by \cite{petroliagis1996}, the authors found good results regarding the predictability of the  P94 case. They found consistency (i.e. no sudden changes) among consecutive forecasts. The global analyses and forecasts we used to drive the Meso-NH simulations benefit from the improvements over the last years in terms of model cycle and data assimilation method. Such improvements further reduce any \textit{jumpiness} as compared to that present in the data of \cite{petroliagis1996}. It follows that any subsequent medium-range (i.e. lead-times $\le$ 72 hours) regional model forecast is as reliable as a short-term (i.e. lead-times $\le$ 24 hours) one.
\newline The IVT maps shown in Figures \ref{fig:AR1} and \ref{fig:AR2} demonstrate that an AR landfall occurred during the P4 case and that it played an important role in controlling the storm-total precipitation in northern Piedmont. Following the scale proposed by \cite{ralph2019scale}, such an AR is in category 2 out of 5. In terms of hazardous impacts, a category 2 corresponds to ``Mostly beneficial, but also hazardous'' \citepay{ralph2019scale}. We stress the fact that this subjective assessment was designed for the west coast of the United States and that, as mentioned in the review article by \cite{gimeno2014atmospheric}, for areas with complex topography, AR landfalls are often associated with large amounts of rainfall. What we can deduce from the presence of an AR during the P94 case, is that it supplied the necessary contribution of moisture able to change the Froude number in the Alps (which is about 0.90 according to \citeay{doswell1998diagnostic}). This conclusion agrees well with the findings in \cite{buzzi1998numerical} and \cite{cassardo2002flood}, who demonstrated that the contribution of moisture was not due to the evaporation of air from the Mediterranean Sea. Furthermore, because ARs involve processes at the meso$\alpha$ and $\beta$ scales, this partially explains why the numerical investigations performed more than 25 years ago (listed in Table \ref{tab:ref}) were quite skilful in predicting the precipitation in the mountainous part of the region. One may argue that diverse algorithms for the detection of ARs take into account not only the IVT intensity and duration (as in \citeay{ralph2019scale}), but also criteria applied on Integrated Water Vapour (IWV) and on the geographical extent of the area where IWV conditions are met. For instance, \cite{ralph2004satellite} and \cite{neiman2008meteorological} used an objective identification algorithm that involves the imposition of three conditions: (i) IWV exceeds 20 $mm$, (ii) wind speeds in the lowest 2 $km$ greater than 12.5 $ms^{-1}$ and (iii) a long ($>$2000 $km$) and narrow ($<$1000 $km$) shape. For the P94 case, the conditions on the IWV and wind values are satisfied, whereas the condition on the extent of the area is not met (namely its length is less than 2000 $km$). Nevertheless, what we want to stress is that the water vapour transported meridionally across the mountainous area of Piedmont was sufficient to change the static stability of the air impinging the reliefs and thus favouring large amounts of rainfall. Indeed, Figure \ref{fig:AR1} confirms the presence of a strong horizontal humidity gradient in the low-level airstreams affecting the Italian peninsula \citepay{rotunno2001mechanisms}. The western part of the flow is moist, whereas the eastern one is drier. This latter is deflected westward (i.e. to the left) over the Po Valley, following the mechanism of the barrier winds (\citeay{buzzi2014heavy}; \citeay{buzzi2020barrier}). Such deflection causes a convergence with the flow over Piedmont, which is very humid. The moist flow is forced to rise over the drier one and upstream of the orographic barrier, reducing dynamically the height of the mountains to surmount. In other words, the lower the orographic barrier (i.e. the value of $h$ in Equation \ref{eq:froude}), the greater the Froude number.
\par The flooding of the Piedmont region in November 1994 is remembered because of its memorable impacts on the territory and the society. It drew the attention of the scientific community due to the atmospheric processes involved, which make this event a testbed to investigate orographic precipitation mechanisms. Past numerical experiments were able to capture most of the features of the event. However, we think that reforecasting past severe weather events by testing new models (e.g. convection-permitting), parametrization (e.g. advanced microphysical schemes) and data (e.g reanalyses) is valuable to better understand the forecasting capabilities of current systems.


%
\begin{acknowledgements}
The Meso-NH model is freely available under the CeCILL license agreement. The author wishes to thank the model's developers and the User Support for their help.
\newline The Copernicus Climate Change Service (C3S) is acknowledged for the ERA5 data, which were used to produce the maps in Figures \ref{fig:AR1} and \ref{fig:AR2}.
\end{acknowledgements}

\section*{Conflict of interest}
The author declares that he has no conflict of interest.

\bibliographystyle{spbasic}      
\bibliography{biblio_valcap}     

\end{document}